# Emergent Noncollinearity and Near-Degenerate Magnetic Superlattices in $AT_6X_6$ Kagome Metals

Weiyi Xia[1,2], Wei-Shen Tee[2], Peter Minch[1], Feng Zhang[1,2], Cai-Zhuang Wang[1,2], and Vladimir Antropov[1]

[1]*Ames National Laboratory, U.S. Department of Energy, Iowa State University, Ames,*

*Iowa 50011, USA*

[2]*Department of Physics and Astronomy, Iowa State University, Ames, Iowa 50011, USA*

## Abstract

Ferromagnetic $AT_6X_6$ Kagome compounds are a popular class of systems in which quantum magnetism with topological features has been observed. These systems allow easy chemical substitution, creating an opportunity to fine-tune their properties. In this paper, we present electronic-structure and magnetic ground-state studies of several $AT_6X_6$ compounds with relatively low magnetic-ground-state stability. We find unusual magnetic orderings, including complex spin-spiral states and the formation of magnetic long-range superstructures. While $LiFe_6Ga_6$ and $TiMn_6Ge_6$ retain collinear AFM ground states with low-energy FM/AFM layer sequences, competing spin-spiral and long-period antiferromagnetic structures in $MgFe_6Ga_6$ and a double-spin-spiral ground state in $TiFe_6Ga_6$ were determined. Magnetism in all these systems appears local, with adiabatic energy profiles suggesting non-Heisenberg long-range interactions, including a strong biquadratic term. In $TiMn_6Ge_6$, we found the conditions for magnetic tunneling. Our results show that, in addition to traditional magnetic topological features in such FM Kagome systems, near-degenerate magnetic superstructures suitable for spintronic switching applications can form naturally. Overall, these systems represent a potentially rich playground for neutron diffraction and spintronics experimental studies.

## Introduction

Noncollinear magnetic order directly modifies transport symmetry, magnon excitations, and spintronic response. Noncollinear antiferromagnets feature sublattice moments oriented in multiple directions at the atomic scale, often yielding compensated or weak net magnetization while retaining magnetic symmetry breaking [1–5]. Depending on the magnetic symmetry, chiral or multipolar spin textures can generate momentum-dependent spin polarization and Berry curvature, giving rise to anomalous Hall and Nernst effects, tunneling magnetoresistance in suitable device geometries, and electrically driven order-parameter dynamics [1–5]. Finite-angle spin textures can also reconstruct electronic states, produce hidden or multipolar order, support magnon spin currents, and stabilize incommensurate magnetic superstructures, skyrmion-like textures, or topological Hall responses in chiral spin states [6–16].

The microscopic origins of noncollinear magnetic order fall into four broad categories. The first is exchange competition. In localized-spin descriptions, this includes frustrated isotropic exchange and competing interactions generated by superexchange, double exchange, long-range oscillatory RKKY coupling, or higher-order and multispin terms [7-9,11,12,17]. In itinerant and first-principles descriptions, the same instability can appear as several nearly degenerate extrema in $J(\mathbf{q})$ or maxima in $\chi(\mathbf{q})$ associated with Fermi-surface nesting [7-9,11]. Different temperature dependences of competing interactions can further drive transitions among collinear, noncollinear, and incommensurate phases. Other categories include geometric frustration imposed by lattice topology; relativistic interactions, including Dzyaloshinskii-Moriya interactions (DMI), anisotropic exchange, and single-ion anisotropy; and electronic or lattice effects such as orbital order, charge order, strain, and magnetoelastic coupling [10,12-14,16]. These mechanisms frequently coexist, and real-space exchange models, first-principles J(q), and susceptibility $\chi(\mathbf{q})$ can provide complementary descriptions of the same noncollinear state.

Among these mechanisms, the layered metallic 1-6-6 Kagome compounds are primarily associated with competition among symmetric interlayer exchange interactions. This $AT_6X_6$ family (where A denotes a spacer cation, T the transition metal on the Kagome sites, and X a p-block element) offers a chemically tunable setting for testing how localized moment formation and itinerant exchange cooperate or compete. The transition-metal Kagome layers carry robust Fe or Mn moments, while the A-site spacer and X-site p-block layers tune electron count, interlayer geometry, p-d hybridization, and the effective exchange kernel without destroying the layered Kagome framework [18–23]. As a result, the relevant magnetic degree of freedom is often not moment formation within a Kagome sheet, but the layer-resolved phase relation between ferromagnetic Kagome sheets. In a metallic layered magnet, long-range itinerant exchange mediated through the same d-derived bands that generate Kagome electronic features can select this phase relation. In well-studied systems such as $YMn_6Sn_6$, collinear density-functional-theory (DFT) comparisons alone do not capture the observed incommensurate ground state; spin-spiral calculations and exchange models are needed to describe the frustrated, long-range interlayer coupling [24,25]. Kagome bands also place Dirac crossings, flat-band-derived features, van Hove singularities, and Berry-curvature responses in the same energy window as magnetic exchange [1,18,26–32]. In $AT_6X_6$ compounds, the metallic electronic structure and the layer-stacking magnetic degree of freedom are therefore coupled through the same transition-metal d states.

In this work, we extend the previous work [20] by studying the stability of collinear and noncollinear magnetic orders in eight Fe- and Mn-based $AT_6X_6$ compounds. Although noncollinear magnetism is established in parts of the broader 166 family, especially in Mn-based systems such as $YMn_6Sn_6$, it has not been systematically examined across representative Fe- and Mn-based $AT_6X_6$ compounds. The collinear calculations confirm well-resolved ferromagnetic (FM) or antiferromagnetic (AFM) ground states for $LiMn_6Ge_6$, $NaMn_6Ge_6$, MgFe6Ge6, and $NaFe_6Ge_6$, as reported in [20]. By contrast, the FM and lowest-energy AFM configurations of $LiFe_6Ga_6$, $TiMn_6Ge_6$, $MgFe_6Ga_6$, and $TiFe_6Ga_6$ differ by no more than 5 meV per transition-metal atom. Most importantly, we predict a competition between finite-angle spin-spiral and long-period AFM orders in $MgFe_6Ga_6$ and a double-spin-spiral noncollinear ground state in $TiFe_6Ga_6$. At the same time, all four near-degenerate compounds retain low-energy FM/AFM layer sequences that preserve the local Fe or Mn moments while changing the net magnetization and stray-field condition. Our results indicate that the emergence of noncollinearity and magnetic superlattices in Kagome 166 systems is a systematic feature arising from complicated magnetic interactions between FM-ordered planes. These interactions are very long-range and depend delicately on chemistry and distance.

## Computational Methods

DFT calculations were performed using the Vienna Ab initio Simulation Package [33,34]. The core-valence electron interactions were described using the projector augmented-wave (PAW) method [35,36], and the exchange-correlation functional was treated within the generalized gradient approximation (GGA) as parameterized by Perdew, Burke, and Ernzerhof (PBE) [37]. The electronic wavefunctions were expanded in a plane-wave basis set with a kinetic energy cutoff of 520 eV. Brillouin-zone integration used a uniform reciprocal-space mesh with a spacing of 0.2 $Å^{-1}$ and the Monkhorst-Pack sampling scheme [38], and partial occupancies were treated using Gaussian smearing with a width of 0.05 eV, and the electronic self-consistency threshold was $10^{-6}$ eV.

An additional $MgFe_6Ga_6$ calculation compared spin-spiral states in a $1 \times 1 \times 1$ unit cell (SC111) with longer-period collinear states in a $1 \times 1 \times 6$ supercell (SC116). In the layer notation used here, AFM2 denotes ↑↑↓↓, AFM3 denotes ↑↑↑↓↓↓, and AFM6 denotes ↑↑↑↑↑↑↓↓↓↓↓↓; each arrow represents one ferromagnetically ordered Fe Kagome layer along the supercell stacking direction. These calculations used the same 520 eV plane-wave cutoff and a k-point spacing of 0.2 $Å^{-1}$, corresponding to $7 \times 4 \times 6$ and $7 \times 4 \times 1$ k-meshes for SC111 and SC116, respectively.

To systematically evaluate the magnetic ground states, we compare total energies across collinear FM and multiple AFM configurations. Compounds with small collinear energy differences, with $|\Delta E| < 10$ meV per transition-metal atom (noted as 10 meV/T, where T = Fe or Mn in this work), where $\Delta E = E(FM) - E_{min}(AFM)$, were selected for noncollinear spin-spiral calculations [39,40]. We performed these calculations by removing symmetry constraints and employing the generalized Bloch theorem, which allows the magnetic moments to rotate by a specific wavevector q from one unit cell to the next [40]. For each calculation, we sampled spin-spiral angles from 0° to 180° in 15° increments, yielding thirteen distinct spiral angles without requiring large commensurate supercells.

To examine whether a single interlayer rotation angle adequately describes the low-energy magnetic configurations, we extended the calculations to double-spin-spiral configurations. The single-spiral angle θ, which describes the rotation between every second Fe Kagome layer, was fixed at 120° for $MgFe_6Ga_6$ and 135° for $TiFe_6Ga_6$. A second angle φ, describing the rotation within each two-layer block, was then varied independently from 0° to 180° in 15° increments. Calculations were performed using both the standard Fe PAW potential and the Fe_pv PAW potential, which includes the Fe 3p semicore states as valence states. For each PAW potential, the energies were referenced to the corresponding 0° single-rotation state and normalized per Fe atom. All other computational settings were kept identical to those used for the corresponding single-spin-spiral calculations.

Magnetic anisotropy energies (MAE) and orbital magnetic moments were calculated by incorporating spin-orbit coupling (SOC) self-consistently. The MAE was determined as the total-energy difference between configurations with the spin quantization axis oriented out of plane and in plane, following the standard first-principles treatment of magnetocrystalline anisotropy as a SOC-driven total-energy difference [41,42].

Additional calculations for $MgFe_6Ga_6$ were performed within the local-density approximation (LDA) using the Ceperley–Alder exchange-correlation functional as parameterized by Perdew and Zunger [43,44], together with the corresponding LDA PAW potentials. The PBE-optimized geometry was retained, and all other computational parameters were unchanged. This fixed-geometry comparison was used to assess the sensitivity of the magnetic-energy ordering to the exchange-correlation treatment from effects associated with structural relaxation.

# Results and Discussion

## Crystal structures and magnetic superlattices in the 166 lattices

The magnetic behavior of the $AT_6X_6$ family is controlled by the coupling between moment-bearing transition-metal Kagome layers and the A/X spacer blocks. The T atoms form the magnetic Kagome network, whereas the spacer blocks set the electron count, interlayer spacing, local coordination, and exchange pathways between adjacent Kagome sheets. This separation of roles is a defining feature of the 166 lattice. Robust local moments reside mainly on the Fe or Mn Kagome sublattice, while the spacer chemistry determines whether neighboring layers prefer ferromagnetic, antiferromagnetic, near-degenerate magnetic superlattices, or finite-angle noncollinear alignment.

As shown in Fig. 1, the compounds considered here are described by two stuffed-CoSn-derived structural prototypes, referred to as Type 1 (Fig. 1a) and Type 2 (Fig. 1b) [19,20]. Type 1 is the hexagonal *P6/mmm* prototype, in which transition-metal Kagome monolayers are stacked along the crystallographic c direction and are separated by high-symmetry spacer blocks. This structure retains a similar interlayer environment and is the dominant prototype for the Ge-containing compounds considered here. Type 2 is the orthorhombic *Immm* prototype. In this structure, A atoms occupy the honeycomb centers of the spacer layers in an ordered single-column pattern, thereby lowering the in-plane symmetry of the spacer block and producing inequivalent interlayer exchange paths.

For each compound, the lattice parameters and internal coordinates were first fully relaxed in the FM state. Static total-energy calculations were then carried out for the competing collinear magnetic-superlattice configurations to identify the lowest-energy collinear reference state and to determine whether further noncollinear calculations were required. For compounds selected for detailed magnetic ground-state analysis, the lowest-energy collinear configuration serves as the structural and magnetic reference for subsequent relaxation and spin-spiral calculations.

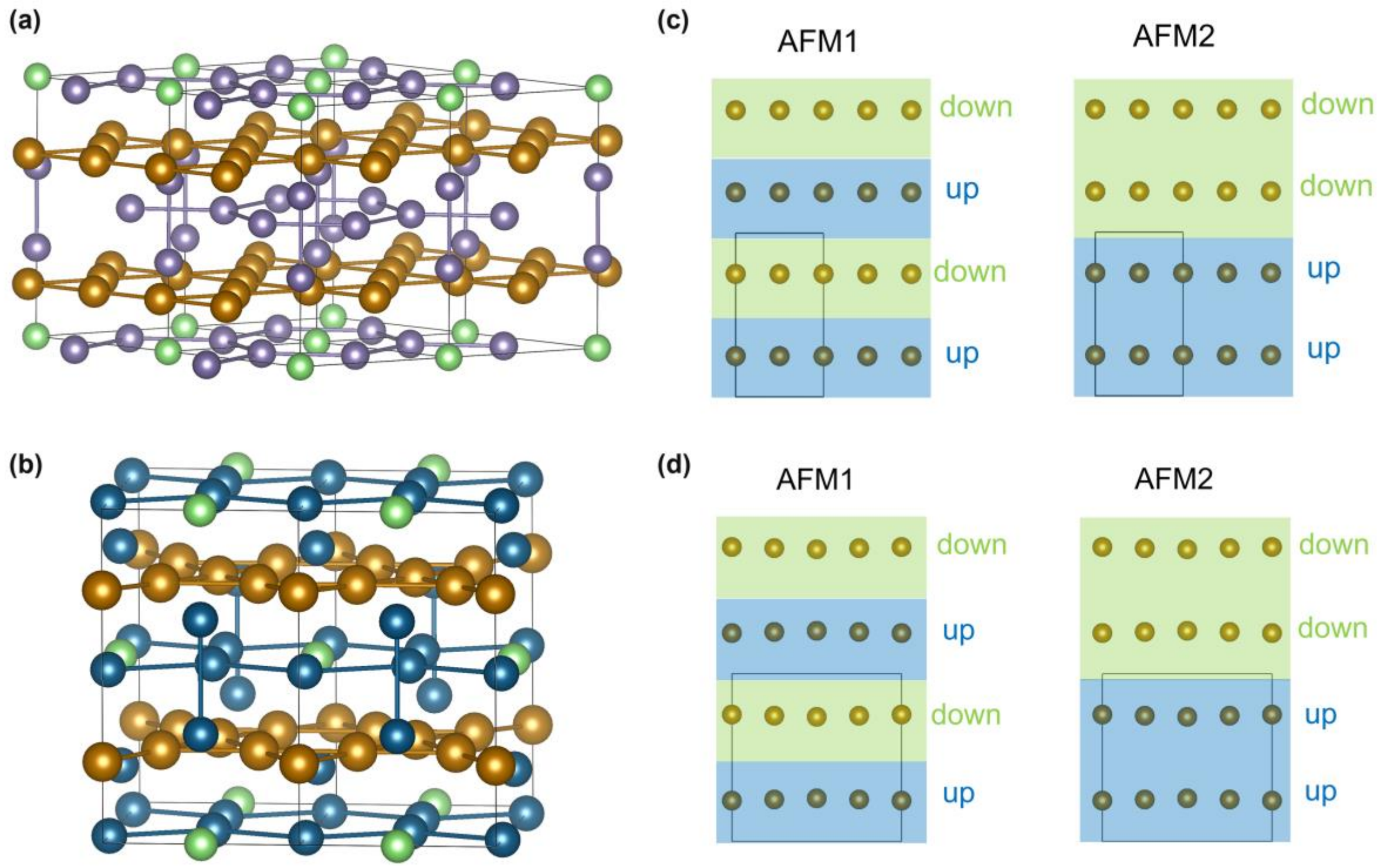

**Figure 1 | Structural motifs and collinear magnetic-superlattice basis for the $AT_6X_6$ screen.** (a) Type-1 P6/mmm structural motif, with Kagome layers stacked along the crystallographic c direction. (b) Type-2 *Immm* structural motif, in which the ordered spacer block lowers the symmetry and makes the crystallographic b axis the out-of-plane layer-stacking direction in the standard CIF setting. (c) Type-1 layer-stacking schematic showing the AFM1 and AFM2 magnetic configurations. (d) Type-2 layer-stacking schematic showing the corresponding AFM1 and AFM2 configurations. Within each transition-metal Kagome layer, the moments are ferromagnetically aligned, and the indicated up/down labels denote the relative layer-resolved moment directions.

The magnetic configurations in Fig. 1 represent layer-stacking variables. We assume strong intralayer FM coupling within each Kagome sheet, consistent with the behavior established in 166 Kagome magnets, where the dominant magnetic competition occurs between ferromagnetic Kagome layers [20,24,25,45]. Multiple AFM configurations were therefore calculated by changing the relative orientation of these ferromagnetic Kagome sheets. AFM1 (AFM2) denotes antiferromagnetic alignment between adjacent transition-metal layers (two-layer blocks of layers); analogous longer-period AFMX configurations can be defined by extending this layer-stacking pattern. These FM, AFM1, AFM2, and longer-period layer sequences are collinear magnetic superlattice states of the 1-6-6 lattice.

We selected the eight compounds from the previous study [20]. $MgFe_6Ge_6$ has been synthesized, whereas the other seven compositions remain predictions. $LiMn_6Ge_6$, $MgFe_6Ge_6$, $NaFe_6Ge_6$, $LiFe_6Ga_6$, $MgFe_6Ga_6$, $TiMn_6Ge_6$, and $TiFe_6Ga_6$ lie on the calculated 0 K convex hull, and $NaMn_6Ge_6$ is 1.9 meV/atom above it [20]. At this energy, it is thermodynamically stable once computational uncertainties are accounted for. Nevertheless, low energy is not sufficient to establish synthesizability, because finite-temperature free energies, competing phases, disorder, and kinetic pathways also affect phase formation [46,47]. Ref. [20] further reports no imaginary phonon branches and satisfaction of the Born stability criteria for $LiMn_6Ge_6$ and $NaMn_6Ge_6$. Taken together, these results indicate that these compounds are strong candidates for synthesis.

Table 1 summarizes the collinear magnetic calculations. Following the convention used in the related $AT_6X_6$ work [20], $\Delta E = E(FM) - E_{min}(AFM)$, where $E_{min}(AFM)$ is the energy of the lowest-energy AFM configuration. Negative $\Delta E$ favors FM order, positive $\Delta E$ favors the lowest-energy AFM configuration, and $|\Delta E|$ gives the FM/AFM energy separation.

**Table 1 | Magnetic energies and magnetic anisotropy of lowest-energy collinear magnetic states of $AT_6X_6$ compounds.** $\Delta E = E(FM) – Emin (AFM)$, and all values are normalized per transition-metal atom. MAE values are calculated using the FM configuration, with units of meV/cell.

| Compound | Type | Lowest energy state | Moment (μB/T) | ΔE (meV/T) | FM MAE (meV/cell) | Anisotropy |
|---|---|---|---|---|---|---|
| $LiMn_6Ge_6$ | 1 | FM | 2.21 | -26 | -0.32 | - |
| $NaMn_6Ge_6$ | 1 | FM | 2.24 | -17 | 0.05 | z |
| $MgFe_6Ge_6$ | 1 | AFM1 | 1.67 | 13 | 1.02 | z |
| $NaFe_6Ge_6$ | 2 | AFM1 | 1.81 | 15 | 2.72 | plane |
| $LiFe_6Ga_6$ | 2 | AFM2 | 2.00 | 5 | 0.84 | plane |
| $MgFe_6Ga_6$ | 2 | FM | 1.93 | -3 | 1.73 | plane |
| $TiMn_6Ge_6$ | 1 | AFM2 | 2.12 | 3 | 0.23 | z |
| $TiFe_6Ga_6$ | 2 | AFM2 | 1.90 | 1 | 3.11 | plane |

We first observe that all eight compounds carry sizable Fe or Mn moments, ranging from 2.12 to 2.24 $\mu_B$ for Mn and from 1.67 to 2.00 $\mu_B$ for Fe, confirming that robust magnetism is intrinsic to the transition-metal Kagome layers. The compounds differ primarily in the stiffness of the relative orientation between neighboring Kagome sheets. In metallic Kagome magnets, this distinction is essential because the same d electrons generate local exchange splitting and mediate longer-range, Fermi-surface-sensitive interlayer exchange.

$LiMn_6Ge_6$, $NaMn_6Ge_6$, $MgFe_6Ge_6$, and $NaFe_6Ge_6$ have $|\Delta E|$ = 13-26 meV/T whereas $LiFe_6Ga_6$, $MgFe_6Ga_6$, $TiMn_6Ge_6$, and $TiFe_6Ga_6$ have $|\Delta E| \leq 5$ meV/T. For the first four compounds, the energy separation between the FM and the lowest-energy AFM configurations is on the order of 10 meV/T or larger; the magnetic ground state can usually be reliably assigned within the collinear configurations considered here. The latter four compounds contain low-energy FM and AFM layer sequences with nearly unchanged local moments but different net magnetization and stray-field conditions.

$LiMn_6Ge_6$ and $NaMn_6Ge_6$ favor FM order, with $\Delta E$ = -26 and -17 meV/T, respectively. $MgFe_6Ge_6$ and $NaFe_6Ge_6$ favor AFM1 order, with $\Delta E$ = 13 and 15 meV/T. A single interlayer alignment dominates the calculated zero-temperature ground states in these four compounds, although the energy differences do not determine their ordering temperatures or domain structures.

The remaining four compounds, $LiFe_6Ga_6$, $MgFe_6Ga_6$, $TiMn_6Ge_6$, and $TiFe_6Ga_6$ have $|\Delta E| \leq 5$ meV/T. In particular, $TiFe_6Ga_6$ has the smallest $|\Delta E|$ of 1 meV/T. These few-meV splittings show that FM and AFM interlayer couplings nearly compensate. The broader $AT_6X_6$ compounds similarly associate near-degenerate FM/AFM energies with long-ranged, higher-order, or non-Heisenberg exchange [20]. Under such exchange frustration, the system may lower its total energy by leaving the collinear 0- or 180- degree spin orientation and forming a spiral, fan-like, or more complex noncollinear texture. This mechanism is well established in $RMn_6Sn_6$-type 166 Kagome magnets and has recently been identified in $LiFe_6Ge_6$ [24,25,45]. The small collinear splittings in $LiFe_6Ga_6$, $MgFe_6Ga_6$, $TiMn_6Ge_6$, and $TiFe_6Ga_6$ thus provide the motivation for the explicit noncollinear calculations discussed below, where spin-spiral noncollinear calculations sample the spin orientation continuously by allowing spins between neighboring layers to rotate by an intermediate angle [39,40,45]. Spiral and fan-like order generated by frustrated interlayer exchange is established in $RMn_6Sn_6$-type magnets and $LiFe_6Ge_6$ [24,25,45]. However, more long ranged Heisenberg or biquadratic terms have been ignored in those studies. The spin-spiral calculations for $LiFe_6Ga_6$, $MgFe_6Ga_6$, $TiMn_6Ge_6$, and $TiFe_6Ga_6$ sample $E(\theta)$ continuously between neighboring Kagome layers [39,40,45], separating near-degenerate collinear superlattices from compounds with lower finite-angle minima. We note that nearly degenerate magnetic states are not a nuisance—they are a signature of rich, tunable, and potentially functional magnetic behavior, but they fundamentally limit both DFT and machine-learning predictive certainty.

The robust Mn-Ge ferromagnets and Fe-Ge antiferromagnets behave as systems in which one interlayer stacking tendency dominates the low-energy magnetic Hamiltonian. The Fe-Ga compounds instead approach a compensated exchange condition, consistent with the broader $AT_6X_6$ trend, in which Ga-containing Fe systems often lie near magnetic degeneracy [20]. Replacing Ge by Ga changes the electron count and *p-d* hybridization, while the A-site species modifies the spacer potential, layer separation, and local Fe coordination. In a metallic system, these substitutions reshape the exchange kernel rather than simply modifying a single local superexchange path. The small $|\Delta E|$ in $MgFe_6Ga_6$ and $TiFe_6Ga_6$ are therefore the first indication that spacer chemistry can place Fe-based 166 compounds near finite-angle noncollinear order.

## Electronic structure of the 166 compounds

We then investigate the electronic structures of these compounds within the lowest-energy collinear states as reference states, as shown in Fig. 2. The transition-metal *d* electrons carry both the local moments and the itinerant interlayer exchange, so the magnetic configuration determines exchange splitting, spin degeneracy, magnetic band folding, SOC-allowed gaps, and orbital character near $E_F$. Recent work shows that FeGe and $MgFe_6Ge_6$ cannot be represented by a single-orbital Kagome model; their dense Fe-derived bands contain several d-orbital sectors coupled to *p* orbitals [30]. Orbital availability, out-of-plane bonding, crystal-field splitting, and *d*-orbital symmetry determines whether Dirac and flat-band-derived features remain identifiable near $E_F$ [31,32].

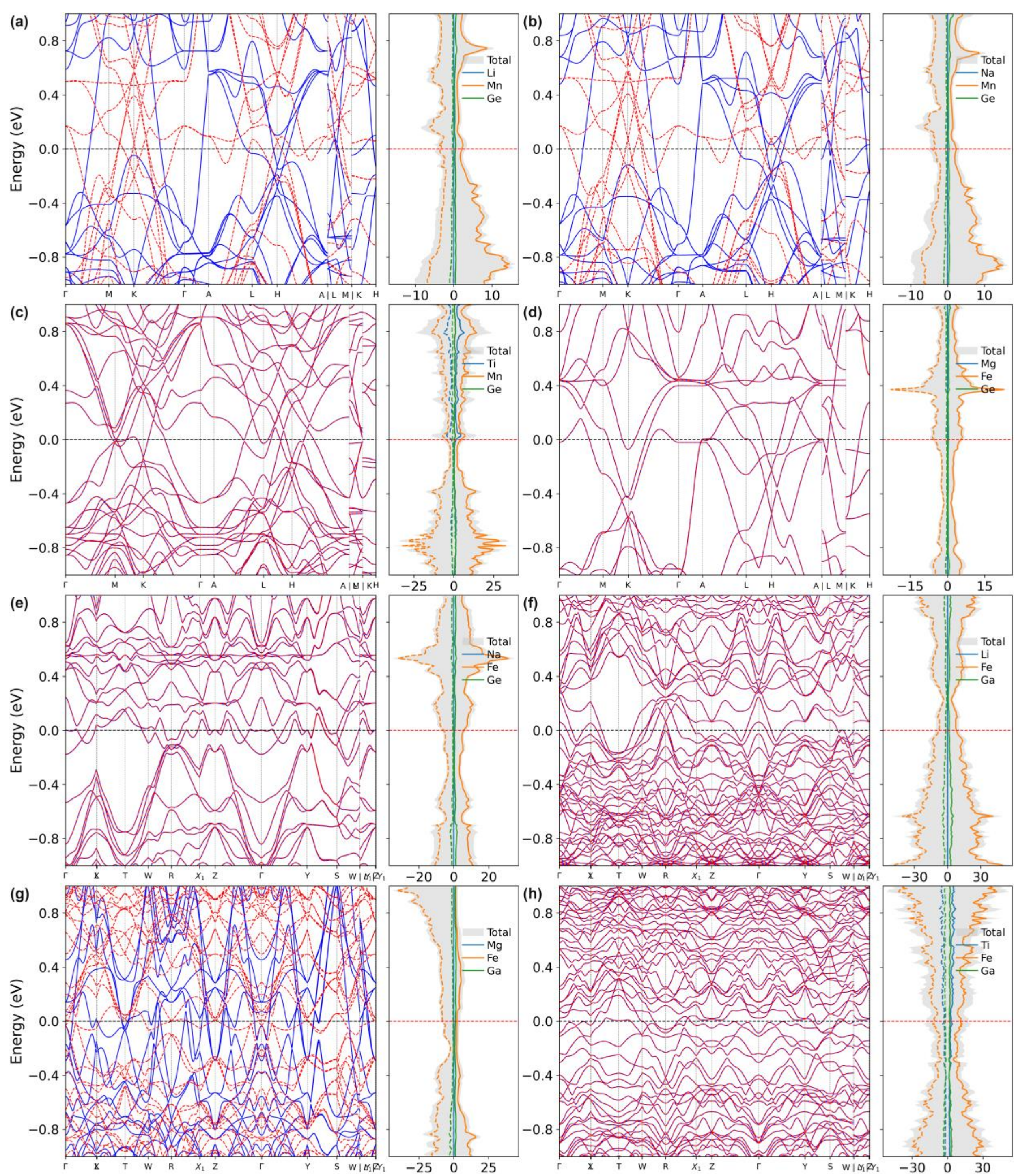

**Figure 2 | Band structures and densities of states for $AT_6X_6$ compounds in the calculated collinear magnetic configurations.** (a) $LiMn_6Ge_6$ in the FM ground state. (b) $NaMn_6Ge_6$ in the FM ground state. (c) $TiMn_6Ge_6$ in the AFM2 lowest-collinear state. (d) $MgFe_6Ge_6$ in the AFM1 ground state. (e) $NaFe_6Ge_6$ in the AFM1 ground state. (f) $LiFe_6Ga_6$ in the AFM2 lowest-collinear state. (g) $MgFe_6Ga_6$ in the FM lowest-collinear reference state. (h) $TiFe_6Ga_6$ in the AFM2 collinear reference state. In the FM panels, blue solid and red dashed bands denote the two spin channels. In the AFM panels, the two spin channels are nearly degenerate and overlap. Gray shading denotes the total DOS; solid curves denote the majority spin and dashed curves the minority spin, with different colors denoting the three elements: blue for A, orange for T, and green for X. The Fermi energy $E_F$ is set to zero.

All eight compounds are metallic within the plotted energy window. The DOS near $E_F$ is dominated by the transition-metal sublattice, whereas the A-site contribution is small. The A atom changes electron count, lattice dimensions, and the spacer potential, but the Fe/Mn d states carry most of the low-energy spectral weight, produce the local moments, and mediate the interlayer exchange that selects collinear or finite-angle alignment.

$LiMn_6Ge_6$ and $NaMn_6Ge_6$ (Fig. 2a,b) contain strongly exchange-split Mn-derived bands crossing $E_F$ in both spin channels and are spin-polarized metals rather than half-metals. Their Li and Na contributions near $E_F$ are small. Replacing Li by the larger Na atoms preserves the almost identical electronic structures. This similarity shows that the Mn-Ge Kagome sheet determines the dominant low-energy electronic structure, while the A site changes band filling, interlayer hopping, and the energies of Kagome-derived features. Previous $AT_6X_6$ calculations on related $LiMn_6Ge_6$ and $NaMn_6Ge_6$ compounds identified minority-spin Dirac-like crossings near K, van Hove features near M or along Γ-M, and flat-band-like Mn-d states above $E_F$ [20]. A similar Dirac point is observed at K for the minority spin slightly above $E_F$.

$TiMn_6Ge_6$ (Fig. 2c) differs qualitatively from the alkali Mn-Ge ferromagnets. Its lowest collinear magnetic state is AFM2, and $\Delta E$ is only about 3 meV/Mn. This change in magnetic ordering relative to $LiMn_6Ge_6$ and $NaMn_6Ge_6$ reflects the different number of valence electrons associated with Ti and the low-lying Ti d states that hybridize with the Mn-Ge-derived manifold. The resulting AFM2 band structure shows that the Dirac point at K shifted below the Fermi level.

The Fe-Ge compounds $MgFe_6Ge_6$ and $NaFe_6Ge_6$ (Fig. 2 d, e) have robust local Fe moments and AFM1 interlayer stacking. The DOS near $E_F$ is dominated by Fe *d* states, with comparatively small contributions from Mg, Na, and Ge in the near-Fermi-level window. The local exchange splitting remains large on each Fe site, while antiferromagnetism folds the spin-polarized Fe bands into a magnetic unit cell with no net ferromagnetic moment. $MgFe_6Ge_6$ in Fig. 2d contains several flat bands: one lies close to $E_F$ along the Γ-A path, and another near 0.4–0.5 eV across several symmetry-line segments. The corresponding Fe-d DOS contains shoulders and peaks in the same energy intervals. These features are consistent with the folded multiorbital Kagome bands reported for FeGe-derived compounds [30].

We identify $NaFe_6Ge_6$ (Fig. 2e) as the AFM1 ground state, with $\Delta E$ = 15 meV/T and a large MAE of 2.72 meV/cell. Its orthorhombic reference structure produces a more folded band representation than the type-1 hexagonal panels. $LiFe_6Ga_6$, $MgFe_6Ga_6$, and $TiFe_6Ga_6$ (Fig. 2f-h) have smaller FM/AFM energy separations than the Fe-Ge AFM1 compounds and dense Fe-derived bands near $E_F$. Replacing Ge by Ga changes valence-electron count and p-d hybridization, while the type-2 *Immm* structure introduces inequivalent interlayer paths. These changes modify the itinerant exchange interactions that couple the Fe Kagome sheets.

The results for $LiFe_6Ga_6$, $MgFe_6Ga_6$, and $TiFe_6Ga_6$ (Fig. 2f-h) show how Ga substitution drives Fe-based 166 chemistry toward a softer magnetic state. Compared with the robust Fe-Ge AFM1 compounds, the Fe-Ga systems have smaller collinear energy separations and a dense set of Fe-derived bands close to $E_F$. Replacing Ge by Ga changes the valence electron count and p-d hybridization, while the type-2 orthorhombic *Immm* geometry introduces inequivalent interlayer paths. In a metallic Fe system, these changes reshape the exchange kernel through the low-energy electronic structure rather than through a single local superexchange path.

$LiFe_6Ga_6$ (Fig. 2f) has $|\Delta E|$ = 5 meV/T, with AFM2 as the lowest collinear state. $TiFe_6Ga_6$ has an even smaller $|\Delta E|$ of 1 meV/T, and favors AFM2 among the collinear configurations, although the double-spin-spiral calculation lowers the energy further. The electronic structure of $TiFe_6Ga_6$ shows a much higher Fe-d density of states, compared with $LiFe_6Ga_6$ and $MgFe_6Ga_6$. Unlike the predominantly ionic Li and Mg spacers, Ti also introduces low-energy *d* states that hybridize with the Fe-Ga bands. The resulting changes in electron filling, orbital hybridization, and interlayer hopping can modify the q-dependent exchange between Fe Kagome layers.

$MgFe_6Ga_6$ differs from the Li and Ti compounds because FM lies 3 meV/T below the lowest-energy AFM configuration. Mg contributes little electronic weight near $E_F$, leaving the low-energy states dominated by the Fe-Ga layers. Replacing Li by Mg increases the nominal valence-electron count by one electron per formula unit and shifts the filling of these metallic Fe-Ga bands. Such a change in filling can reverse the sign of the Fermi-surface-sensitive interlayer exchange. More detailed calculations and discussion are shown below.

Isotropic exchange interactions select the relative orientation of neighboring Kagome layers, whereas SOC energy determines the orientation of the moments relative to the crystal lattice. For robust collinear compounds, calculated anisotropy can often be related to the spin orientation of the ground state once the final sign convention is fixed. For some compounds, however, the magnetic ground states are not FM, as discussed later. Their FM reference MAE values should therefore be interpreted as measures of intrinsic orbital hardness and SOC sensitivity, rather than as anisotropy energies of the true ground states.

From a spintronics perspective, the energy scale for changing the relative alignment of neighboring Kagome layers should be distinguished from the SOC energy that orients the moments relative to the crystal lattice. $TiFe_6Ga_6$ has the largest calculated FM-reference MAE, 3.11 meV/cell (1.41 $MJ/m^3$), followed by $NaFe_6Ge_6$ at 2.72 meV/cell (1.20 $MJ/m^3$); both significantly exceed the approximately 0.40 $MJ/m^3$ value of hcp Co [41]. Together with the few-meV separations among competing interlayer configurations, these values indicate that some Fe-based compounds combine weak energetic selection among layer sequences with appreciable SOC anisotropy.

## Spin spiral calculations

To determine the ground states of $LiFe_6Ga_6$, $MgFe_6Ga_6$, $TiMn_6Ge_6$, and $TiFe_6Ga_6$, we performed spin-spiral calculations using the generalized Bloch theorem. In all four cases, the competing collinear states have energy differences within 5 meV between FM and AFM2. AFM2 has the lowest energy among the possible AFM configurations in these four compounds, indicating strong FM coupling between nearest-neighbor layers but a preferred AFM coupling between two-layer blocks. We therefore set up spin spirals along the z-axis, with rotations between every two layers from 0° (FM) to 180° (AFM2).

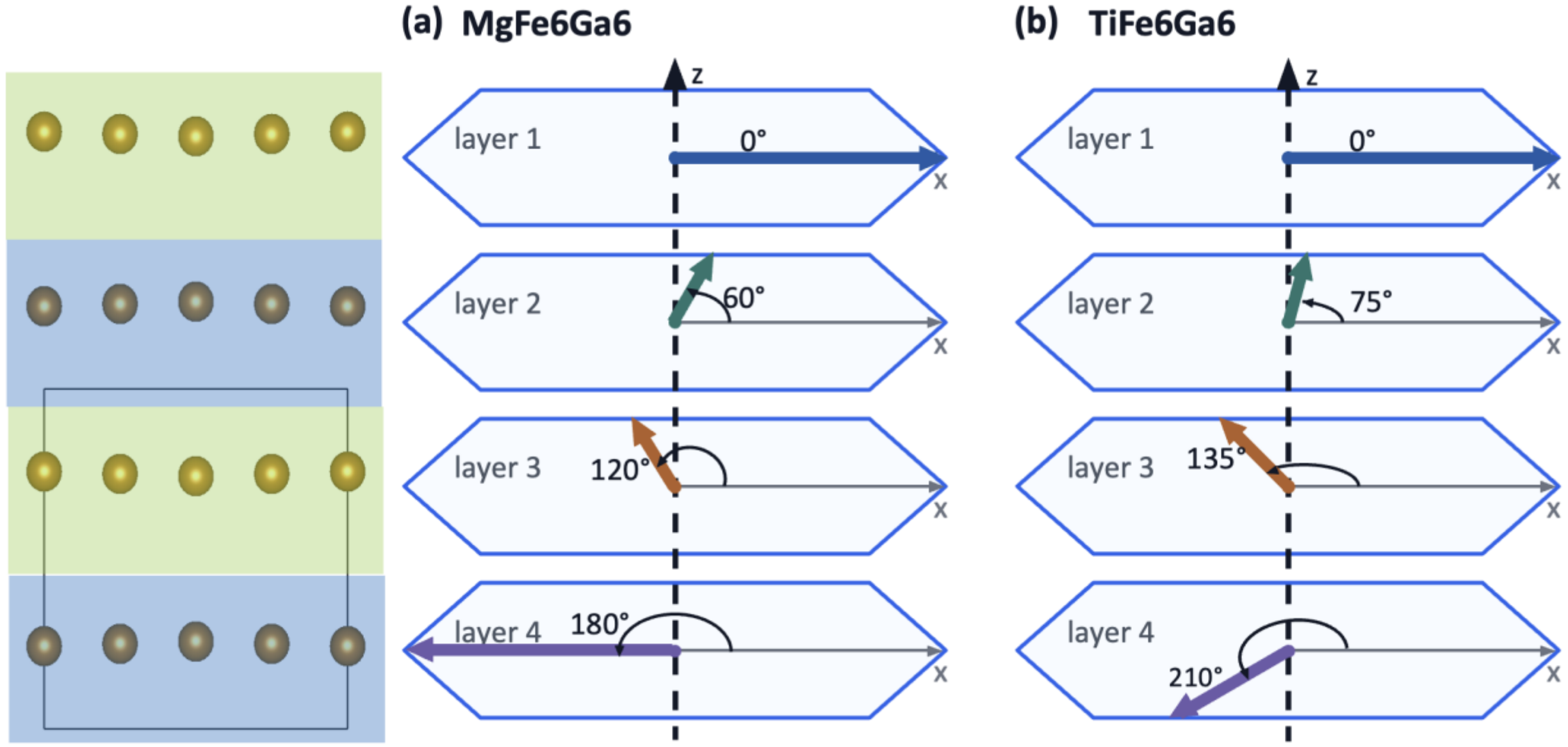

**Figure 3 | Layer-resolved schematic of representative spin-spiral states in the Fe-Ga compounds.** The left panel gives the layer-stacking view. (a) $MgFe_6Ga_6$ is shown using the symmetric 60°/60° partition of the 120° spin spiral between every second Fe layer. (b) $TiFe_6Ga_6$ is represented by adjacent rotations of approximately 75°, 60°, and 75° across four layers, giving a 135° rotation between every second transition-metal layer.

If magnetic behavior were governed by one dominant bilinear interlayer exchange interaction, the layer-rotation energy would have a single global minimum, giving a minimum at θ = 0° (FM) or 180° (corresponding to AFM2 for the near-degenerate cases considered here). More generally, for a uniform single-q spiral, the energy may be written schematically as $E(\theta) = E_0 + J_1 \cos(\theta) + J_2 \cos^2(\theta) + ...$, where $J_1$ represents bilinear exchange, and $J_2$ represents a biquadratic term. Bilinear exchange between layers separated by two stacking intervals produces the same cos(2θ) dependence as biquadratic exchange. Distinguishing these contributions requires additional nonuniform magnetic configurations, such as the double-spin-spiral states considered below [7,9,11,20,33,41,48].

Figure 3 shows representative layer-resolved spin rotations for $MgFe_6Ga_6$ and $TiFe_6Ga_6$. For $MgFe_6Ga_6$, θ = 120° denotes the rotation between every second Fe Kagome layer. The schematic uses the symmetric 60°/60° partition between adjacent layers; the results below show that this partition is not uniquely selected. The $TiFe_6Ga_6$ configuration contains adjacent-layer rotations of approximately 75°, 60°, and 75°, yielding a 135° rotation between every second transition-metal layer.

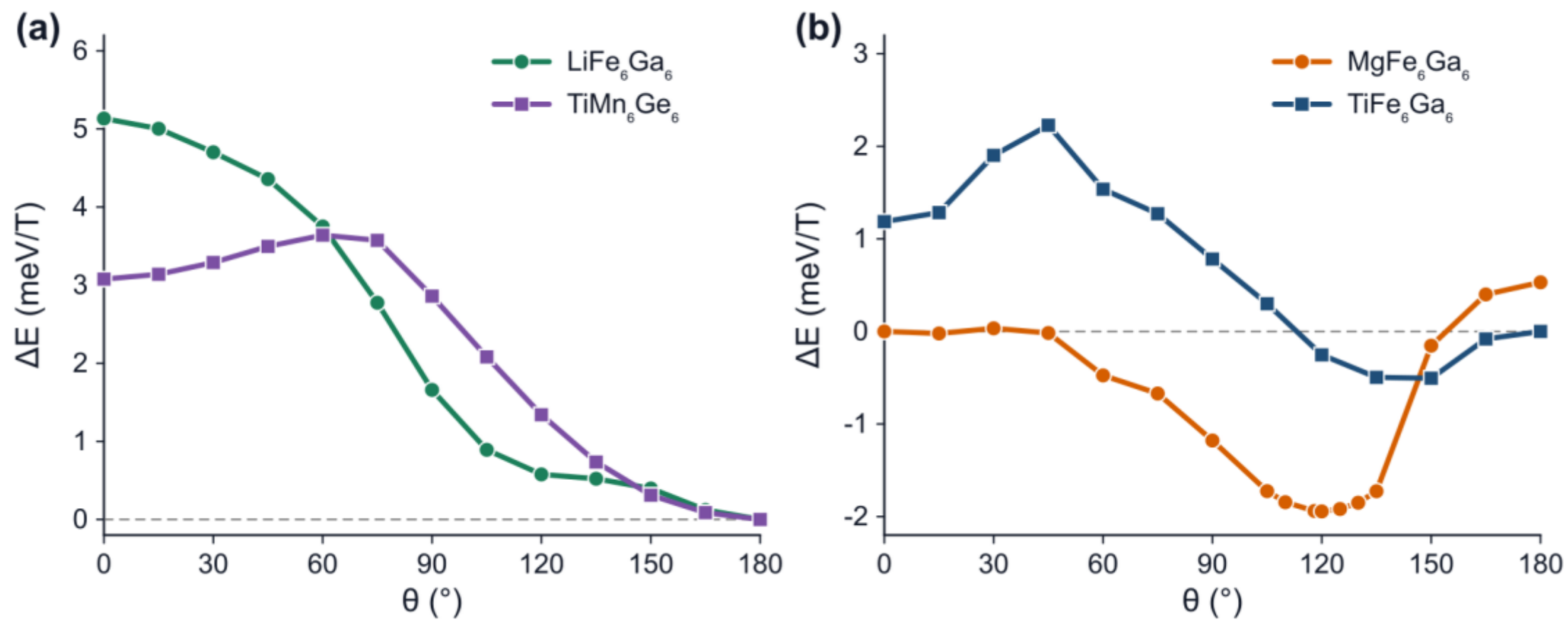

**Figure 4 | Single-spiral energy-angle calculations for the near-degenerate compounds in Table 1.** (a) $LiFe_6Ga_6$ and $TiMn_6Ge_6$. (b) $MgFe_6Ga_6$ and $TiFe_6Ga_6$. $LiFe_6Ga_6$ and $TiMn_6Ge_6$ have minima at the AFM2 (180°) state, whereas $MgFe_6Ga_6$ and $TiFe_6Ga_6$ have finite-angle minima below the FM/AFM2 reference states. Energies are normalized per transition-metal atom, and we use the lowest collinear states as reference energies.

Fig. 4 shows the results of our noncollinear spin-spiral calculations, with total energy as a function of the spin-spiral angle between every two transition-metal layers along the layer-stacking direction. $LiFe_6Ga_6$ and $TiMn_6Ge_6$, although both have |ΔE| within 5 meV, do not develop a lower-energy minimum at an intermediate angle, as shown in Fig. 4a. The energy of $LiFe_6Ga_6$ decreases toward the AFM2 endpoint in a manner consistent with predominantly bilinear Heisenberg exchange, and AFM2 remains its magnetic ground state. $TiMn_6Ge_6$ instead exhibits a pronounced maximum near θ = 60°, indicating a substantial departure from a bilinear Heisenberg form and the presence of higher-order, biquadratic-type contributions. The resulting E(θ) profile makes $TiMn_6Ge_6$ a magnetic analog of a two-well potential, for which magnetic tunneling may be considered [49,50]. Strong biquadratic interactions have also been established in iron pnictides, where they arise from itinerant magnetic interactions and substantial longitudinal moment fluctuations [51,52]. The robust local moments in the present compounds instead favor a biquadratic interaction within a localized-spin Hamiltonian. Earlier descriptions of spiral order in ferromagnetic $RMn_6Sn_6$ Kagome layers predominantly used bilinear Heisenberg exchange [53].

These results demonstrate that small |ΔE| identifies the coexistence of FM/AFM magnetic superlattices without establishing a noncollinear ground state. Such a small energy difference between the FM and AFM states could be useful in spintronic switching devices.

For $MgFe_6Ga_6$, we found the energy minimum between 115° and 130° (Fig. 4b). Additional calculations further confirmed that the energy minimum occurs close to 120° between two transition-metal layers. The resulting energy minimum is lowered by about 1.94 meV/T relative to the FM configuration. The energy curve clearly shows a biquadratic-like interaction, distinct from the bilinear Heisenberg behavior in $LiFe_6Ga_6$ and $TiMn_6Ge_6$ (Fig. 4a), indicating higher-order exchange contributions [7,9,11,20,48].

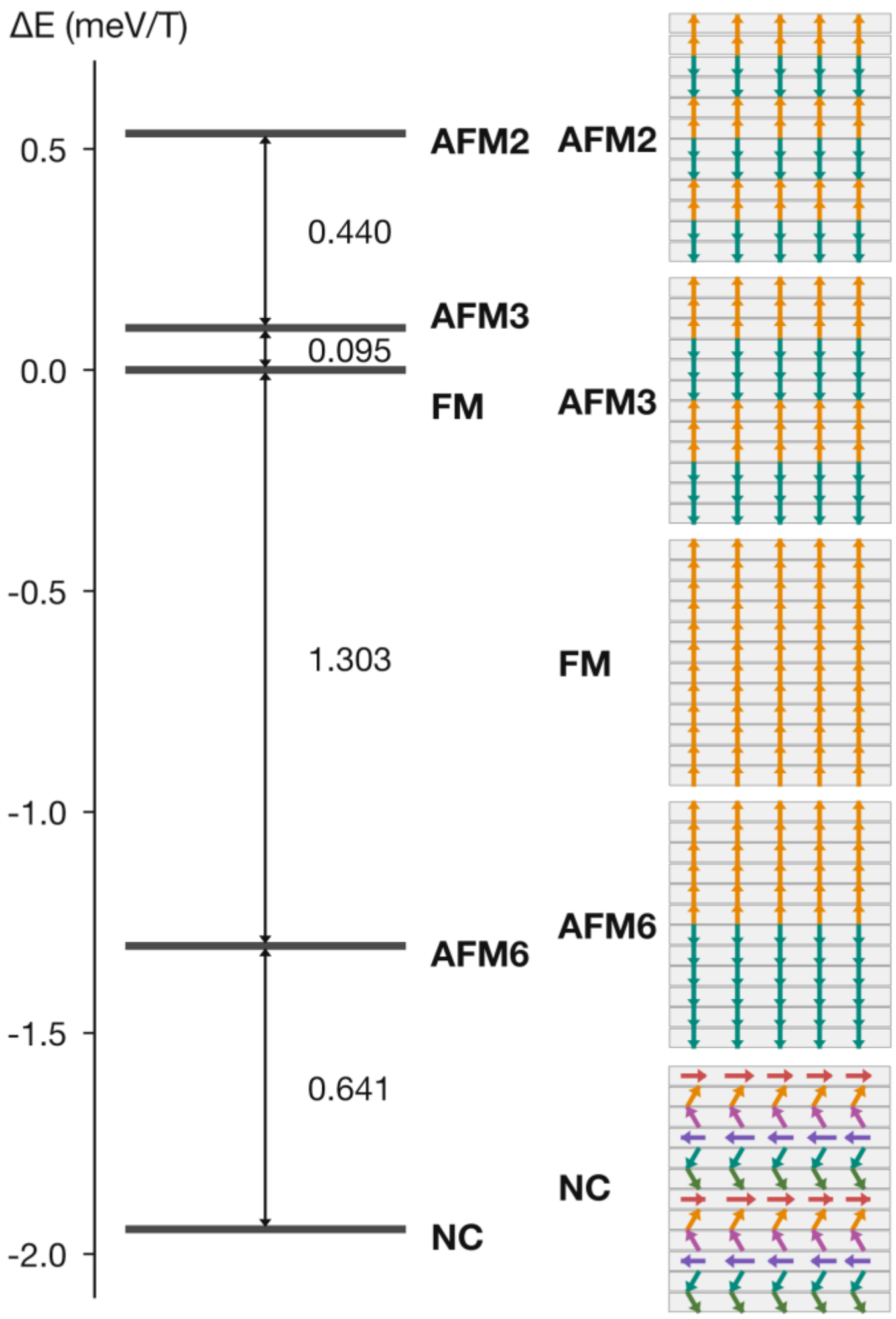

**Figure 5 | Finite-angle and longer-period collinear magnetic states of $MgFe_6Ga_6$.** Relative-energy levels and layer-resolved magnetic configurations of $MgFe_6Ga_6$. Energies are normalized per Fe atom. AFM2, AFM3, FM, AFM6, and the θ = 120° noncollinear (NC) state are arranged from higher to lower relative energy. Each bar represents one ferromagnetically ordered Fe Kagome layer; orange and green arrows denote opposite collinear moment directions. The schematic for the spin-spiral NC state uses the arrows with 60° symmetric partition of 120° spin rotation between every two layers, and the same color is reused when the same angle repeats.

Fig. 5 compares the finite-angle minimum directly with longer-period collinear magnetic superstructures using a 1 × 1 × 6 supercell. Among the collinear states, AFM6, with a complex configuration of 6 consecutive layers spin up and another 6 consecutive layers spin down, is lowest at −1.30 meV/T relative to FM. On the other hand, the spin-spiral calculation for the primitive cell reaches −1.94 meV/T at θ = 120°. This energy difference is within the reliable energetic resolution of the present calculations and may depend on k-point density, choice of pseudopotential, plane-wave cutoff, and self-consistency thresholds. Therefore, $MgFe_6Ga_6$ is predicted to have competing long-ranged antiferromagnetic configurations and finite-angle noncollinear states. The existence of several competing magnetic configurations with very similar energies suggests an unusual magnetic adiabatic energy landscape, in which multiple energy minima of collinear and noncollinear states coexist already at T = 0 K. One can expect strong competition of these states at finite temperatures to form the magnetic ground state.

We also obtained LDA magnetic stability results with PBE-optimized geometry, as described in Methods section. The LDA calculations produce a qualitatively different magnetic-energy ordering from PBE, as shown in Table 2. For $MgFe_6Ga_6$, PBE favors the θ = 120° spin-spiral state, followed by AFM6, whereas LDA favors AFM2 state, with energy of 0.603 meV/T below FM and approximately 0.5 meV/T below both AFM6 and the θ = 120° state. The Fe moments are 1.93 μB/Fe with PBE and 1.80 μB/Fe with LDA. The difference in magnetic stability is therefore not associated with a substantial change in the moment magnitude. Instead, it demonstrates strong sensitivity of the competing interlayer exchange interactions to the exchange-correlation potential. Such disagreement can be resolved only experimentally. This functional sensitivity may indicate that long-range magnetic interactions of potential-exchange origin contribute to the magnetic ordering, which may be important in these materials.

**Table 2 | Magnetic energies of lowest-energy collinear and noncollinear magnetic states of $MgFe_6Ga_6$ compound calculated with PBE/LDA functional.** Energies are referenced to FM and normalized per transition-metal atom.

| Magnetic state | PBE (meV/T) | LDA (meV/T) |
|---|---|---|
| FM | 0 | 0 |
| AFM2 | 0.535 | -0.603 |
| AFM3 | 0.095 | 0.292 |
| AFM6 | -1.303 | -0.135 |
| NC (θ = 120°) | -1.944 | -0.119 |

The results for $TiFe_6Ga_6$ also show a biquadratic-like exchange interaction. Its collinear states (FM and AFM2) have |ΔE| of only about 1 meV/T, and the spin-spiral calculations identify the spin spiral between 135° and 150° (Fig. 4b) as the relevant low-energy noncollinear states. However, this energy is only 0.505 meV/T below the AFM2 state. This difference is not large enough to identify this spin-spiral state as the true magnetic ground state; additional magnetic configurations are required to separate long-range bilinear exchange from higher-order itinerant terms [48], as discussed below.

The centrosymmetric crystallographic structures do not support a uniform DMI as the leading origin of these turn angles. Competing symmetric exchange and DMI-driven spirals produce different interactions and excitation spectra [12,25,54]. In $MgFe_6Ga_6$ and $TiFe_6Ga_6$, spacer chemistry changes the compensation among itinerant exchange paths that can depend on Fermi-surface structure, nesting, and long-range oscillatory coupling [8,9,20].

Our calculations identify that both $LiFe_6Ga_6$ and $TiMn_6Ge_6$ have AFM2 order as the ground state. $MgFe_6Ga_6$ contains competing spin-spiral and AFM6 states that degenerate within the present numerical accuracy, whereas $TiFe_6Ga_6$ has a double spin-spiral ground state with energy well below the collinear states. We further tested whether the spin-spiral calculations in $MgFe_6Ga_6$ and $TiFe_6Ga_6$ are robust against the Fe pseudopotentials, since the energies are only on order of meV/T, and the treatment of Fe 3*p* semicore states can shift small magnetic energy differences. Figure 6 compares the total energies of spin-spiral calculations obtained with the standard Fe potential and the Fe_pv potential, where Fe_pv treats the 3*p* semicore states as valence states.

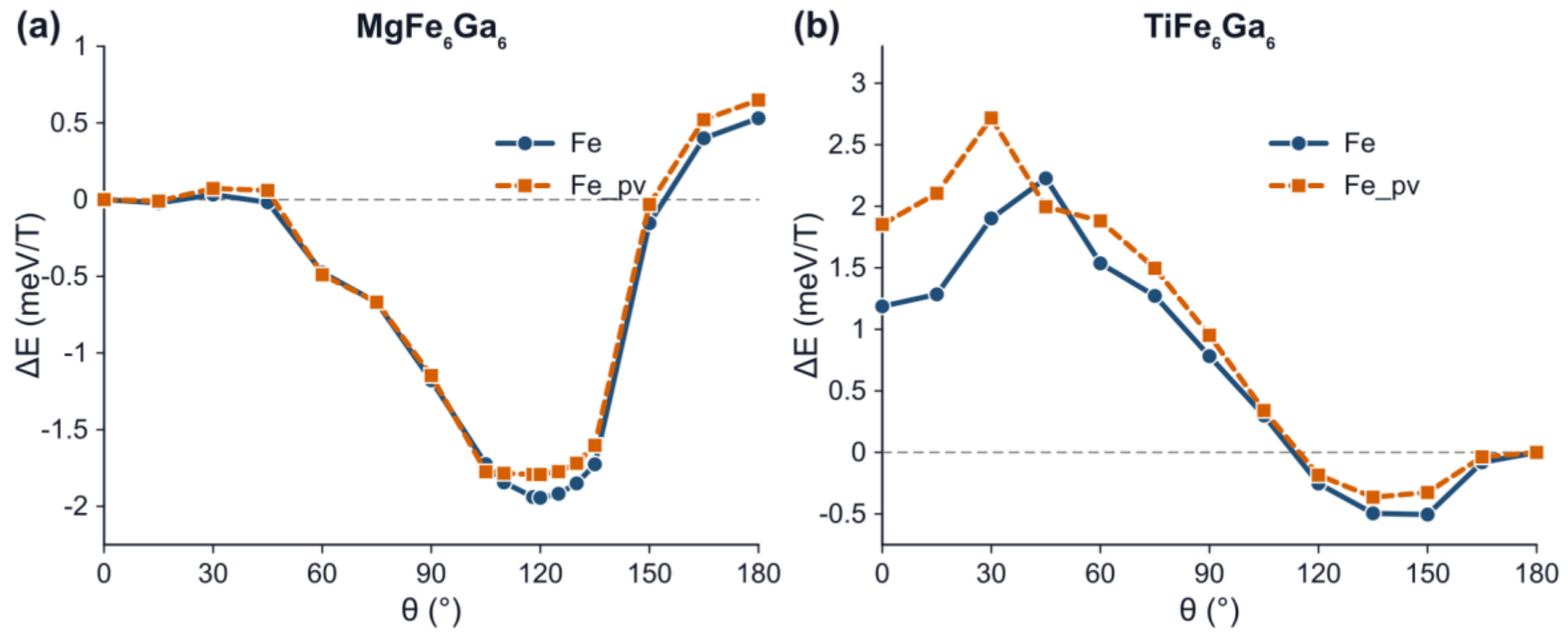

**Figure 6 | Fe-pseudopotential consistency checks for the single-spiral minima in $MgFe_6Ga_6$ and $TiFe_6Ga_6$.** (a) $MgFe_6Ga_6$ calculations with the standard Fe (blue) and Fe_pv (orange) potentials. (b) $TiFe_6Ga_6$ calculations with the standard Fe (blue) and Fe_pv (orange) potentials. With both potentials, $MgFe_6Ga_6$ retains a minimum near 120°, and $TiFe_6Ga_6$ retains a minimum near 135°.

For $MgFe_6Ga_6$, the standard Fe potential gives a single-spiral minimum of spiral angle around 120°, while the Fe_pv results have a minimum between 105° and 130°. The energy is -1.944 meV/T relative to the FM reference with standard Fe potential and -1.793 meV/T with Fe_pv potential. For $TiFe_6Ga_6$, the standard Fe potential and Fe_pv potential both place the lowest energy state at around 135°, with energy minima of 0.505 and 0.364 meV/T below their respective AFM2 references. With such a small energy difference, we cannot conclude that this noncollinear state with a spin spiral at 135° is truly the magnetic ground state.

Because the type-2 *Immm* structure contains inequivalent interlayer environments, a single uniform spiral angle may not span all relevant magnetic degrees of freedom. We therefore extended the analysis for $MgFe_6Ga_6$ and $TiFe_6Ga_6$ to double-spin-spiral configurations, allowing two adjacent interlayer rotations to vary independently around the single-spiral minima. We also performed calculations using both Fe and Fe_pv potentials (Fig. 7).

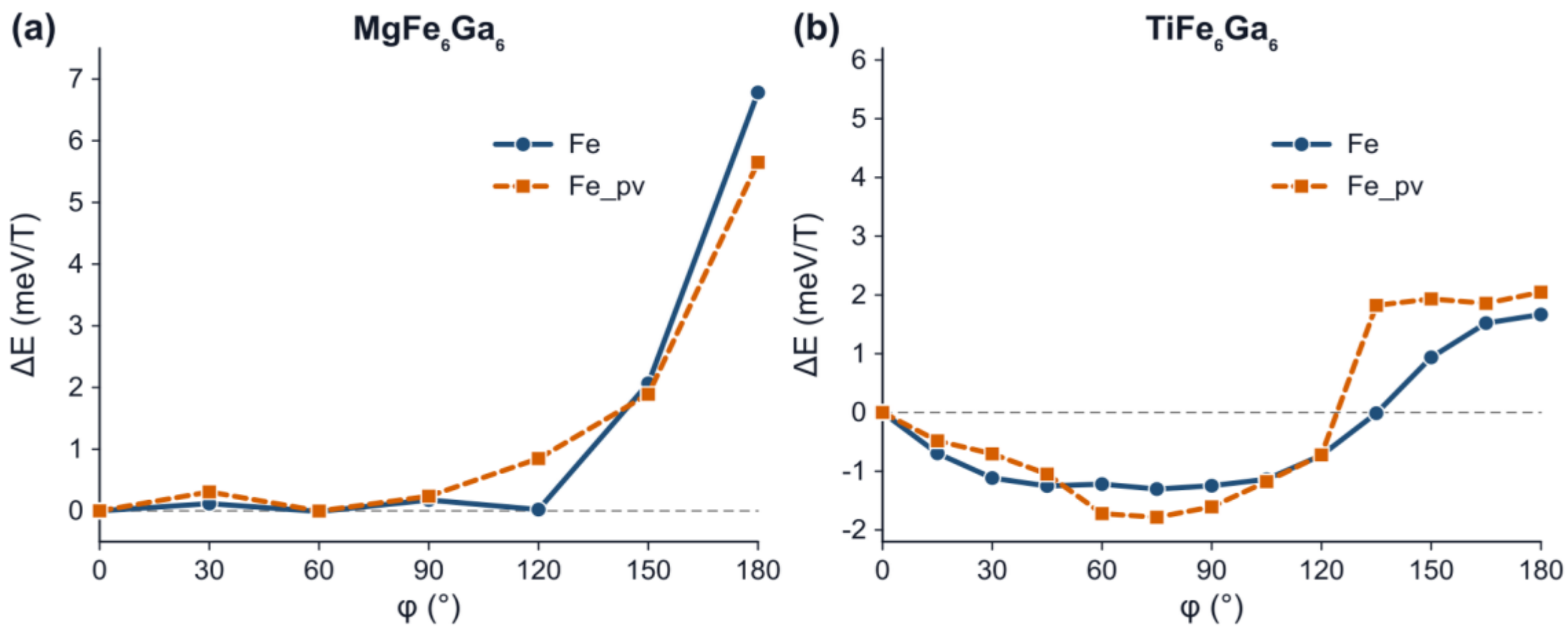

**Figure 7 | Double-spin-spiral calculations with two Fe pseudopotentials for $MgFe_6Ga_6$ and $TiFe_6Ga_6$.** (a) $MgFe_6Ga_6$ calculations starting from the fixed θ = 120° single-spiral state with the standard Fe (blue) and Fe_pv (orange) potentials. (b) $TiFe_6Ga_6$ calculations starting from the fixed θ = 135° single-spiral state with the standard Fe (blue) and Fe_pv (orange) potentials. The varied angle φ describes the rotation within each two-layer block, and the complementary adjacent-layer rotation is θ − φ. Each curve is referenced to its corresponding φ = 0° state and normalized per Fe atom.

For $MgFe_6Ga_6$, the double-spin-spiral calculations around the θ = 120° single-spiral state produce a local minimum at φ = 60° and at φ = 120° with both Fe potentials, with the energy lowering relative to φ = 0° less than 0.01 meV/T, indicating almost no energetic preference for redistributing the adjacent interlayer spin angle of the two-layer 120° spin spiral. The detailed φ dependence differs between the two potentials, particularly at φ = 120°, but neither potential stabilizes an additional double-spiral modulation by a significant energy. The robust conclusion is that the noncollinear state, with a spin spiral of θ = 120° for every two FM layers, lies approximately 1.9 meV/T below the collinear FM state, while the spin rotation between the adjacent layers is not uniquely determined. PBE results with both potentials place this state 0.5–0.6 meV/T below AFM6, whereas LDA makes the two states nearly degenerate, while place AFM2 approximately 0.5 meV/T below both states. The ordering between the spin spiral states, AFM2 and the long-period AFM6 superstructure therefore remains unresolved.

$TiFe_6Ga_6$ retains a lowest-energy double-spin-spiral state with both potentials. The minimum combines a two-layer rotation near θ = 135° with an independent adjacent-layer rotation near φ = 75°. The standard Fe potential gives approximately 1.30 meV/T below the 135° single spiral state and 1.80 meV/T below the collinear AFM2 reference. Calculations with the Fe_pv potential give similar results, with slightly different energies of 1.78 meV/T below the single spiral state and 2.14 meV/T below AFM2. This two-angle configuration is the lowest state among those sampled for $TiFe_6Ga_6$.

The double-spiral result also clarifies why a nearest-neighbor Heisenberg description is insufficient. Together, these calculations reported here provide evidence for the first time in the $AT_6X_6$ series, for a genuine biquadratic contribution beyond an ad hoc nearest-neighbor Heisenberg model with frustrated exchange. Longer-range bilinear and itinerant multispin interactions can coexist with this contribution [7,9-11,48]. These long-range interactions generate the collinear and noncollinear magnetic superstructures found here.

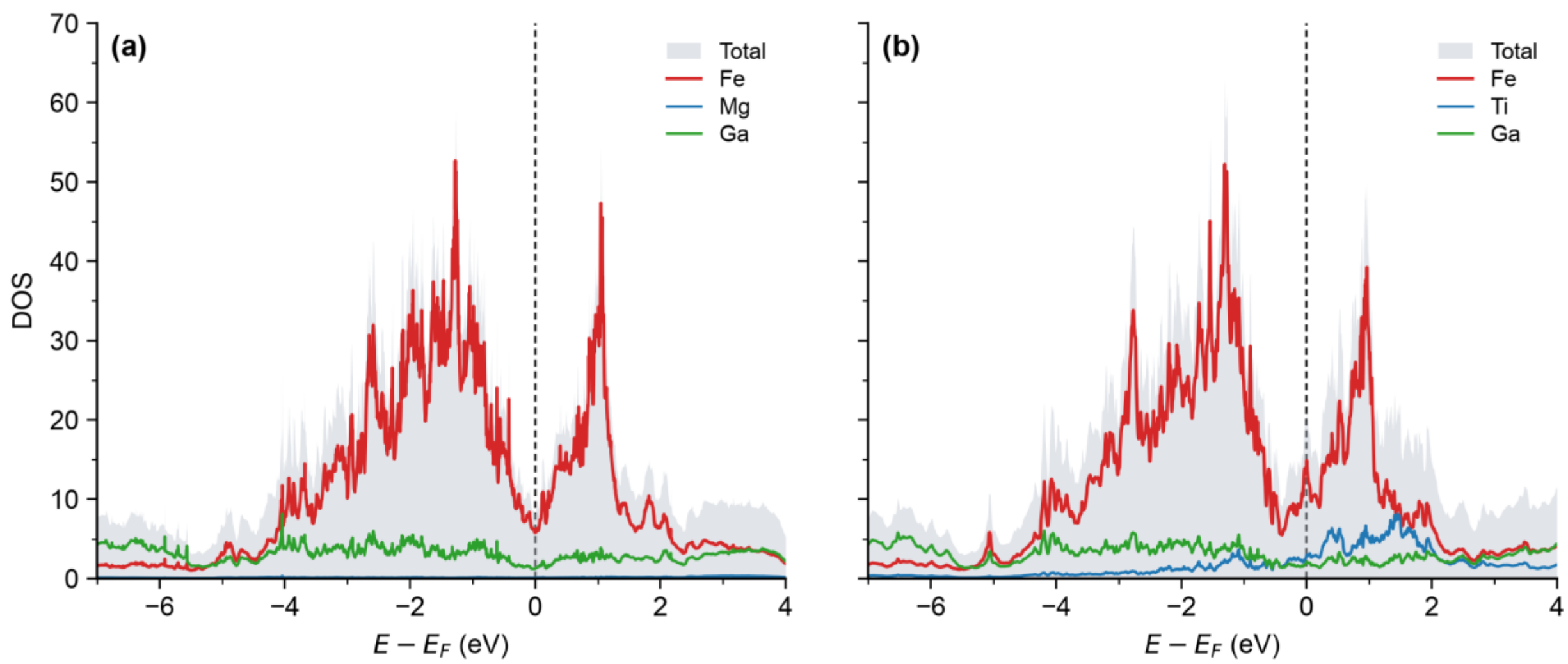

**Figure 8 | Full-range total and element-projected densities of states of the calculated noncollinear configurations** (a) $MgFe_6Ga_6$ and (b) $TiFe_6Ga_6$. Gray shading denotes the total DOS, while the red, blue, and green curves denote the Fe, Mg/Ti, and Ga contributions, respectively. The Fermi energy $E_F$ is set to zero. DOS values are reported in states $eV^{-1}$ per 26-atom unit cell.

The electronic structures of the noncollinear $MgFe_6Ga_6$ and $TiFe_6Ga_6$ configurations are compared in Fig. 8 and Fig. 9 (enlarged view). Both configurations remain metallic, with Fe *d* states dominating the energy window near $E_F$ in both compounds. While we observe a DOS dip near $E_F$ for $MgFe_6Ga_6$ noncollinear state with $\theta = 120°$, we identify a DOS peak near $E_F$ and an appreciable Ti-d contribution in $TiFe_6Ga_6$, indicating potential magnetic instability in this compound. A higher DOS at $E_F$ means this material experimentally will be much more sensitive to perturbations such as pressure, strain, chemical substitution, and temperature.

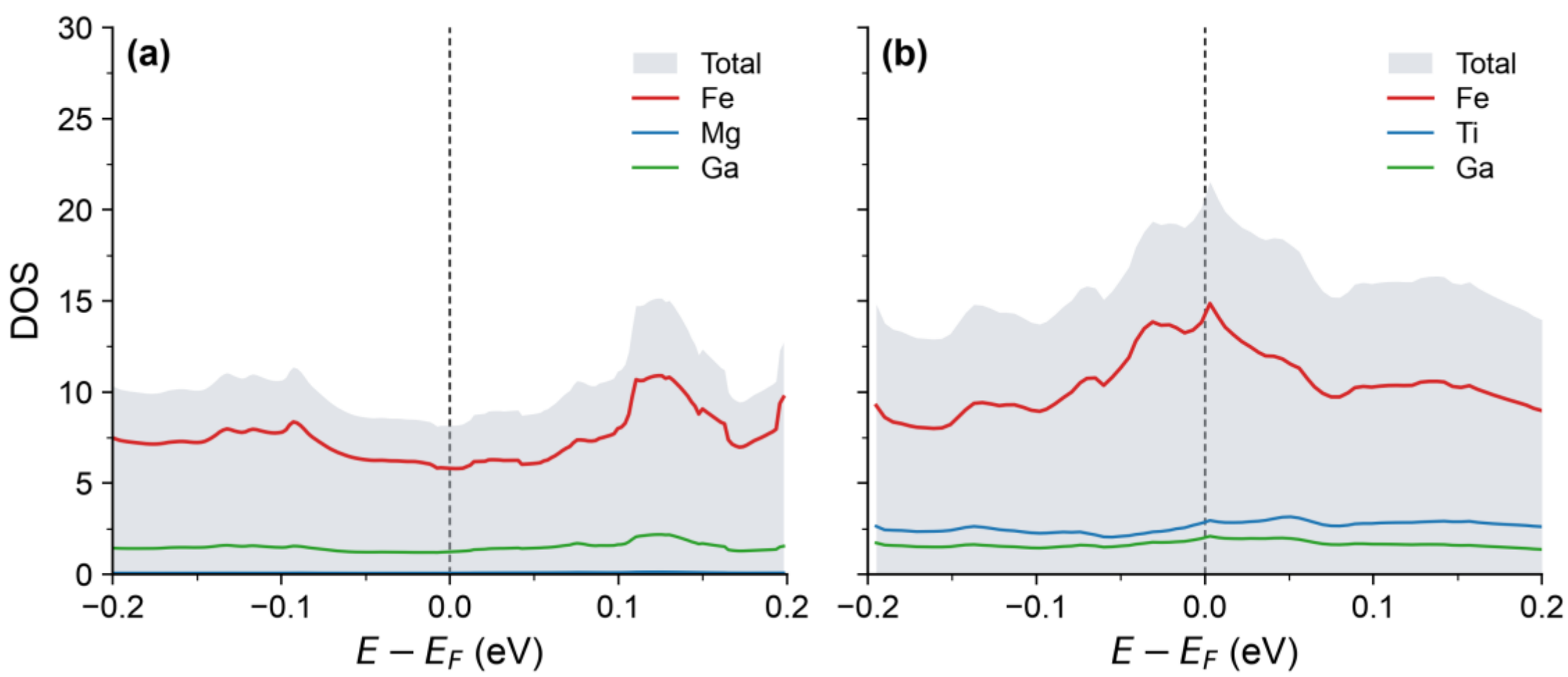

**Figure 9 |** Enlarged view of projected density of states of noncollinear (a) $MgFe_6Ga_6$ and (b) $TiFe_6Ga_6$. Gray shading denotes the total DOS, while the red, blue, and green curves denote the Fe, Mg/Ti, and Ga contributions, respectively. The Fermi energy $E_F$ is set to zero. DOS values are reported in states $eV^{-1}$ per 26-atom unit cell.

The near degeneracy of the FM, AFM, and finite-angle configurations indicates weak wave-vector selectivity of the magnetic energy along the interlayer direction, consistent with a shallow effective exchange spectrum $J(\mathbf{q})$. The enlarged projected DOS of $TiFe_6Ga_6$ in Fig. 9 (b) reveals a pronounced Fe-$d$ peak at $E_F$. These features are consistent with the Fermi level lying close to a van Hove singularity. In an itinerant Kagome metal, such near-Fermi-level saddle points and weakly dispersive states can produce broad or multiple maxima in the bare susceptibility $\chi_0(\mathbf{q})$, reducing the energetic selectivity among ordering wave vectors and thereby promoting competition among FM, AFM, and finite-angle states [7–9,18,20,30–32]. This provides a plausible microscopic origin for the competing FM, AFM, and finite-angle states.

The meV-scale energy separations ($|\Delta E| \leq 5$ meV/T) among FM-like, AFM-like, and noncollinear layer sequences in $LiFe_6Ga_6$, $MgFe_6Ga_6$, $TiMn_6Ge_6$, and $TiFe_6Ga_6$ identify a low-energy layer-alignment degree of freedom relevant to magnetic switching studies. In synthetic antiferromagnets, interlayer exchange enables field- or current-driven reconfiguration of magnetic layers [17,55,56], while parallel and antiparallel layer alignments can produce giant magnetoresistance (GMR) or tunneling magnetoresistance (TMR) in appropriate transport geometries [57,58]. The $AT_6X_6$ compounds provide a chemically ordered setting in which several layer configurations occur within a few meV/T, without requiring an artificially deposited spacer multilayer. The calculated energetic proximity motivates calculations and measurements of switching fields, activation barriers, spin-orbit torque efficiencies, and state-dependent transport. In addition, $MgFe_6Ga_6$ and $TiFe_6Ga_6$ host low-energy noncollinear states, for which magnetic-symmetry and spin-resolved transport analyses are needed to determine whether spin-split antiferromagnetic or current-induced torque responses are realized [1–4,59].

## Summary

In summary, we have investigated Fe and Mn-based $AT_6X_6$ Kagome compounds by combining collinear total-energy comparisons, spin-spiral calculations, magnetic-anisotropy calculations, and electronic-structure analysis. The Fe and Mn Kagome layers support robust local moments, whereas the principal variation among the compounds is the coupling between successive magnetic layers. $LiMn_6Ge_6$ and $NaMn_6Ge_6$ favor ferromagnetic order, while $MgFe_6Ge_6$ and $NaFe_6Ge_6$ favor collinear antiferromagnetic order. Their $|\Delta E|$ of 13–26 meV/T gives a well-resolved collinear magnetic ground state. The other four compounds, $LiFe_6Ga_6$, $TiMn_6Ge_6$, $MgFe_6Ga_6$, and $TiFe_6Ga_6$, lie in a different regime, with competing FM and AFM layer stackings with $|\Delta E| \leq 5$ meV. All eight compounds are metallic, with the states near the Fermi level dominated by Fe or Mn $d$-orbital states. The Fe compounds exhibit larger SOC total-energy differences. The same near degeneracy that limits the determination of the magnetic ground state is also the source of magnetic tunability. It leaves several layer sequences and ordering vectors sufficiently close in energy to be reorganized by chemical substitution, temperature, magnetic field, or electrical current.

The spin-spiral calculations show that a small ΔE in these systems is a necessary but not sufficient condition for the appearance of a noncollinear ground state. $LiFe_6Ga_6$ and $TiMn_6Ge_6$ retain AFM2 as magnetic ground states. $MgFe_6Ga_6$ and $TiFe_6Ga_6$ behave differently. In $MgFe_6Ga_6$, the noncollinear states lie approximately 1.94 meV/T below the FM state, with spin spirals of 120° between every two Fe layers and rotation between each adjacent layer not uniquely determined; a longer-period AFM6 configuration, lies 0.64 meV/T above the 120° state with PBE. LDA instead favors AFM2, which lies 0.603 meV/T below FM and approximately 0.5 meV/T below AFM6 and the θ = 120° state. Therefore, $MgFe_6Ga_6$ is considered an unresolved magnetic ground state of magnetic instability, with competing collinear long-layer stacking AFM ordering and spin-spiral states. $TiFe_6Ga_6$ is a noncollinear ground-state candidate among the compounds examined. Spin-spiral calculations show that the magnetic ground state is characterized by a rotation of approximately 135° between every second transition-metal layer and an independent rotation of approximately 75° between adjacent layers. This double-spin-spiral ground state lies approximately 1.8 meV/T below the collinear AFM2 state and 1.3 meV/T below the 135° single-spiral state. Notably, the noncollinear $TiFe_6Ga_6$ state exhibits a pronounced Fe-$d$ DOS peak at $E_F$, together with dense, weakly dispersive bands in its immediate vicinity. This electronic structure is consistent with proximity to a van Hove singularity and provides a plausible itinerant origin for the weak wave-vector selectivity and competition among FM, AFM, and finite-angle magnetic orders.

These results extend finite-angle magnetism in the 166 family beyond the established Mn-based compounds. In $MgFe_6Ga_6$ and $TiFe_6Ga_6$, changing the spacer and $p$-block chemistry brings several interlayer exchange channels onto the same meV energy scale, allowing long-period collinear and finite-angle states to compete. The spin spirals in these centrosymmetric compounds are consistent with competing long-range symmetric exchange and higher-order itinerant contributions rather than with a simple nearest-neighbor Heisenberg interaction. The multiple low-energy collinear and noncollinear states form new magnetic superstructures and flat or multi-well magnetic-energy surfaces that provide static conditions relevant to magnetic reconfiguration and tunneling.

The proposed magnetic structures can be tested directly. Neutron diffraction or resonant X-ray magnetic scattering can distinguish the AFM2, AFM6, and finite-angle configurations of $MgFe_6Ga_6$ and determine whether $TiFe_6Ga_6$ exhibits the proposed double-spiral modulation [24,45,60]. Magnetization and susceptibility measurements can identify a collinear ordered phase followed by a lower-temperature spin reorientation, as observed in $LiFe_6Ge_6$ [45]. Mössbauer spectroscopy would provide an important test of whether the Fe moment remains large upon entering the noncollinear state, following its use in related Li-Fe-Ge compounds [60]. Transport measurements should be interpreted only after the magnetic structure has been determined, because collinear and finite-angle states possess different magnetic symmetries and can therefore produce qualitatively different Berry-curvature responses.

Taking together, the calculations confirm the robust collinear magnetic assignments of the Mn-Ge and Fe-Ge compounds, refine the ground-state descriptions of the four near-degenerate systems, and predict new finite-angle magnetic states in the Fe-Ga compounds. Chemical control of the spacer and $p$-block layers tunes the long-range interlayer exchange from well-resolved collinear order to near-degenerate FM/AFM layer sequences and finite-angle noncollinear magnetism. These outcomes motivate complementary spintronic studies: the noncollinear states require magnetic-symmetry and transport analyses to determine their transverse and current-induced responses, whereas the near-degenerate FM/AFM sequences provide states with different net magnetizations and stray-field conditions. This layer-alignment degree of freedom is analogous to that exploited in synthetic antiferromagnets and magnetic superlattices, where interlayer exchange supports field- or current-driven reconfiguration and magnetoresistance [17,55-59]. In the $AT_6X_6$ compounds, the calculated energy proximity identifies candidates for such switching behavior, but the switching fields, activation barriers, torque efficiencies, and state-dependent transport remain to be established. Together with previously reported thermodynamic and dynamic stability of these compounds, we call for synthesis, neutron diffraction or resonant X-ray magnetic scattering, followed by field- and current-dependent magnetization and magnetotransport measurements.

## Acknowledgments

Work at Ames National Laboratory was supported by the U.S. Department of Energy (DOE), Office of Science, Basic Energy Sciences, Materials Sciences and Engineering Division, including a grant of computer time at the National Energy Research Scientific Computing Center (NERSC), Berkeley, CA. Ames National Laboratory is operated for the U.S. DOE by Iowa State University under Contract No. DE-AC02-07CH11358.